\documentclass[12pt,preprint]{aastex}
\pdfoutput=1
\usepackage{graphicx}
\usepackage{hyperref}
\hypersetup{colorlinks=true,linkcolor=red,citecolor=blue,urlcolor=cyan}
\usepackage{pdflscape}

\slugcomment{Technical Report \& User's Guide}

\begin{document}


\title{The GLENDAMA Database}


\author{Luis J. Goicoechea, Vyacheslav N. Shalyapin\altaffilmark{\star} and 
Rodrigo Gil-Merino}
\affil{Universidad de Cantabria, Santander, Spain}


\altaffiltext{$\star$}{permanent address: Institute for Radiophysics and 
Electronics, Kharkov, Ukraine}


\begin{abstract}
This is the first version (v1) of the Gravitational LENses and DArk MAtter 
(GLENDAMA) database accessible at
\begin{quote}
\url{http://grupos.unican.es/glendama/database}
\end{quote}
The new database contains more than 6000 ready-to-use (processed) astronomical 
frames corresponding to 15 objects that fall into three classes: (1) lensed QSO 
(8 objects), (2) binary QSO (3 objects), and (3) accretion-dominated radio-loud 
QSO (4 objects). Data are also divided into two categories: freely available 
and available upon request. The second category includes observations related 
to our yet unpublished analyses. Although this v1 of the GLENDAMA archive 
incorporates an X-ray monitoring campaign for a lensed QSO in 2010, the rest of 
frames (imaging, polarimetry and spectroscopy) were taken with NUV, visible and 
NIR facilities over the period 1999$-$2014. The monitorings and follow-up 
observations of lensed QSOs are key tools for discussing the accretion flow in 
distant QSOs, the redshift and structure of intervening (lensing) galaxies, and 
the physical properties of the Universe as a whole.
\end{abstract}





\section{Target objects and facilities} \label{objfac}

At present, the Gravitational LENses and DArk MAtter 
(\href{http://grupos.unican.es/glendama/}{GLENDAMA}) project at the {\it 
Universidad de Cantabria} (UC) is 
conducting a programme of observation of eight lensed QSOs with facilities at 
the European Northern Observatory (ENO). This consists of an optical monitoring 
by using several instruments on a robotic 2m telescope, as well as deep imaging 
and spectroscopy with 2$-$10m telescopes. We are also involved in the searching 
and identification of new lens systems. This will allow us to enlarge the 
current list of lensed QSOs and have a few virgin systems for subsequent 
analyses. Our project is aimed to the simultaneous follow-up of a reduced 
sample of target objects ($\leq$ 10), and in its previous phases, we observed 
some binary and accretion-dominated radio-loud QSOs. All objects in the 
\href{http://grupos.unican.es/glendama/database}{GLENDAMA database} are 
described in Table~\ref{tbl-1}.

The project started in 1998, before the robotic revolution, and we used many 
facilities throughout the period 1999$-$2014. Here, we comment on the different 
telescopes, instruments and additional details (filters, spectral arms, grisms,
gratings, etc) corresponding to observations that are included in the GLENDAMA 
archive. Regarding the ENO facilities:
\begin{itemize}
\item Gran Telescopio CANARIAS (\href{http://www.gtc.iac.es/}{GTC}) 
	\begin{itemize}
	\item Optical System for Imaging and low-Intermediate-Resolution 
	Integrated Spectroscopy (OSIRIS): long-slit spectroscopy (grisms R300R,
	R500B and R500R)
	\end{itemize}
\item IAC80 Telescope 
(\href{http://www.iac.es/telescopes/pages/es/inicio/telescopios/iac80.php}{IAC80}) 
	\begin{itemize}
	\item optical CCD camera: Johnson-Bessell ($BVRI$) filters\footnote{Alex 
	Oscoz provided us with frames taken with the IAC80,  within  the  
	framework of an IAC-UC coordinated project.}
	\end{itemize}
\item Isaac Newton Telescope 
	(\href{http://www.ing.iac.es/astronomy/telescopes/int/}{INT}) 
	\begin{itemize}
	\item Intermediate Dispersion Spectrograph (IDS): long-slit spectroscopy
	(grating R300V)
	\end{itemize}
\item Liverpool Telescope (\href{http://telescope.livjm.ac.uk/}{LT}) 
	\begin{itemize}
	\item RATCam optical CCD camera: Sloan ($griz$) filters
	\item IOO optical CCD camera: Sloan ($gri$) filters
	\item Fibre-fed RObotic Dual-beam Optical Spectrograph (FRODOSpec): blue
	and red arms (low resolution gratings)
	\item RINGO optical polarimeter: V+R filter (old configuration) and 3 CCD
	cameras (new configuration). These cameras are called B (blue), G (green)
	and R (red)
	\end{itemize}
\item Nordic Optical Telescope (\href{http://www.not.iac.es/}{NOT}) 
	\begin{itemize}
	\item Andalucia Faint Object Spectrograph and Camera (ALFOSC): long-slit 
	spectroscopy (grisms \#7 and \#14) and imaging (filters \#76 $\equiv$ 
	Bessel $R$ and \#12 $\equiv$ interference $i$) 
	\item Stand-by CCD camera (StanCam): Bessell ($VR$) 
	filters\footnote{These 
	frames are part of the Gravitational LensES International Time Project 
	(GLITP), which was led by the UC and other international institutions.}
	\end{itemize}
\item STELLA 1 Telescope 
	(\href{http://www.aip.de/en/research/facilities/stella/instruments}{STELLA}) 
	\begin{itemize}
	\item Wide Field STELLA Imaging Photometer (WiFSIP): Johnson-Cousins 
	($UBV$) and Sloan ($gr$) filters
	\end{itemize}
\item Telescopio Nazionale Galileo (\href{http://www.tng.iac.es/}{TNG}) 
	\begin{itemize}
	\item Near Infrared Camera Spectrometer (NICS): $JHK$ filters
	\end{itemize}
\item William Herschel Telescope
	(\href{http://www.ing.iac.es/Astronomy/telescopes/wht/}{WHT}) 
	\begin{itemize}
	\item Intermediate dispersion Spectrograph and Imaging System (ISIS): 
	long-slit spectroscopy in the blue and red arms (R300B and R316R 
	gratings)
	\end{itemize}
\end{itemize}
In 2010, we also performed space-based observations with 
\href{http://chandra.harvard.edu/}{{\it Chandra}} and 
\href{http://swift.gsfc.nasa.gov/}{{\it Swift}}:
\begin{itemize}
\item Chandra High Resolution Mirror Assembly (HRMA) 
	\begin{itemize} 
	\item Advanced CCD Imaging Spectrometer (ACIS): S3 (0.1$-$10 keV) 
	\end{itemize}
\item UltraViolet and Optical Telescope (UVOT) 
	\begin{itemize} 
	\item Microchannel-plate Intensified CCD (MIC): filter $U$  
	\end{itemize}
\end{itemize}

\section{Observations and their status} \label{obssta}

Our current observations of QSOs were mainly made with the LT in the context of 
a long-term monitoring programme of gravitational lens systems. All these LT 
observations, specific software tools (photometric pipelines, framework for 
accurate reduction of FRODOSpec data, etc) and associated products (light 
curves, calibrated spectra, etc) are being made available to the community in 
the shortest possible time. In addition, the
\href{http://grupos.unican.es/glendama/database}{GLENDAMA database} 
incorporates valuable non-LT information to make a total of more than 6000 
processed frames of 15 objects over the period 1999$-$2014. These frames and 
their status are shown in Table~\ref{tbl-2}. The key idea is to complete a rich 
photometric, spectroscopic and polarimetric archive for a sample of lensed QSOs 
in the northern hemisphere throughout the first decades of the current century. 
Such a final archive will allow astronomers to improve knowledge on the 
structure of distant active galactic nuclei, the mass distribution in galaxies 
at different redshifts and cosmological parameters. 

We note that the LT is a unique facility for photometric, polarimetric and 
spectroscopic monitoring campaigns of lensed QSOs. However, taking into account 
the spatial resolution (pixel size) of the polarimeter (RINGO) and the two 
current spectrographs (FRODOSpec and SPRAT; this last long-slit spectrograph 
has been made available to general users this year), we only consider 
spectropolarimetric follow-up observations of lens systems with angular size 
larger than or similar to 2$^{\prime\prime}$ (see Table~\ref{tbl-1}). Thus, we
are tracking the evolution of broad-band fluxes for all systems, whereas we are 
obtaining additional spectroscopic and/or polarimetric data of QSO B0957+561, 
SDSS J1001+5027, SDSS J1339+1310, SDSS J1515+1511 and QSO B2237+0305. Our main
spectropolarimetric targets for the LT are QSO B0957+561, SDSS J1001+5027 and 
SDSS J1515+1511, since the other two lensed QSOs have either very faint images 
(SDSS J1339+1310) or very crowded components (QSO B2237+0305). These two 
systems are useful to probe the limits/performance of FRODOSpec and RINGO.

All frames in the GLENDAMA archive are ready-to-use because they were processed 
by standard techniques (sometimes as part of specific pipelines) or more 
sophisticated reduction procedures. For example, the 
\href{http://telescope.livjm.ac.uk/}{LT website} presents pipelines for 
RATCam, IOO, FRODOSpec and RINGO, which contain basic instrumental reductions. 
We offer outputs from the pipelines for RATCam, IOO and RINGO, without any 
extra processing. Thus, potential users should carefully consider whether 
supplementary 
tasks are required, e.g., cosmic ray cleaning, bad pixel mask or defringing. We
do not offer outputs from the L2 pipeline for the 2D spectrograph FRODOSpec
(12$\times$12 square lenslets bonded to 144 optical fibres), but 
multi-extension FITS files, each consisting of four extensions: [0] $\equiv$ L1 
(output from the CCD processing pipeline L1, which performs bias subtraction, 
overscan trimming and CCD flat fielding), [1] $\equiv$ RSS (144 row-stacked
wavelength-calibrated spectra from the L2LENS reduction tool), [2] $\equiv$ 
CUBE (spectral data cube giving the 2D flux in the 12$\times$12 spatial array 
at each wavelength pixel) and [3] $\equiv$ COLCUBE (datacube collapsed over its 
entire wavelength range). 
In the [1$-$2] extensions, the
\href{http://grupos.unican.es/glendama/LQLM_tools.htm}{L2LENS software}
produces the sky-unsubtracted and flux-uncalibrated spectra of interest. The 
main differences between the standard L2 pipeline \citep{Bar12} and the new 
L2LENS tool are described in \citet{Sha14a}. Multi-extension FITS files from L1 
and L2 are also available upon request. These and many other datasets are 
put in a secondary datastore, which is not managed by the software of the 
GLENDAMA archive (see below).

The GTC/OSIRIS, INT/IDS, NOT/ALFOSC and WHT/ISIS long-slit spectroscopy was 
processed using standard methods of bias subtraction, trimming, flat-fielding, 
cosmic-ray rejection, sky subtraction and wavelength calibration. The reduction 
steps of the NOT/ StanCam frames included bias subtraction and flat-fielding 
using sky flats, while the combined frames for deep-imaging observations with 
NOT/ALFOSC were obtained in a standard way. We also applied standard reduction 
procedures to the IAC80 original data, although only the final $VR$ datasets 
contain WCS information in the FITS headers. These most relevant data were also
corrected for cosmic-ray hits on the CCD. The TNG/NICS NIR frames were 
processed with the 
\href{http://www.arcetri.astro.it/~filippo/snap/}{SNAP software}, and 
different types of instrumental reductions are applied to STELLA, {\it Swift} 
and {\it Chandra} observations before data are made available to users. After 
passing quality control, the raw frames from the STELLA robotic observatory are
bias corrected and flat fielded, and WCS details are written to FITZ headers.
The space-based observatories also perform specific processing tasks that are 
outlined in dedicated websites. 

Apart from the database in Table~\ref{tbl-2}, we have also prepared a 
complementary datastore including other materials: raw frames, short-time 
adquisition frames for spectroscopic programmes, exposures for wavelength 
calibrations, etc. Outputs from the L2 pipeline for FRODOSpec and many (but not 
all) raw data from the GLITP optical monitoring of QSO B2237+0305 
\citep{Alc02,Sha02} were put there. We want to recover all GLITP raw frames of 
QSO B2237+0305 (several are not yet localized) for later reduce them and put 
the final products in our public archive. Some datasets in the secondary 
storage are available upon request, and unfortunately, a few monitoring 
programmes in the period 1999$-$2014 could not yet be assembled in neither 
disk-based storage \citep{Col03,Ull06}. 
 


\section{Web user interface} \label{wui}

\begin{figure}
\includegraphics[width=1.3\textwidth]{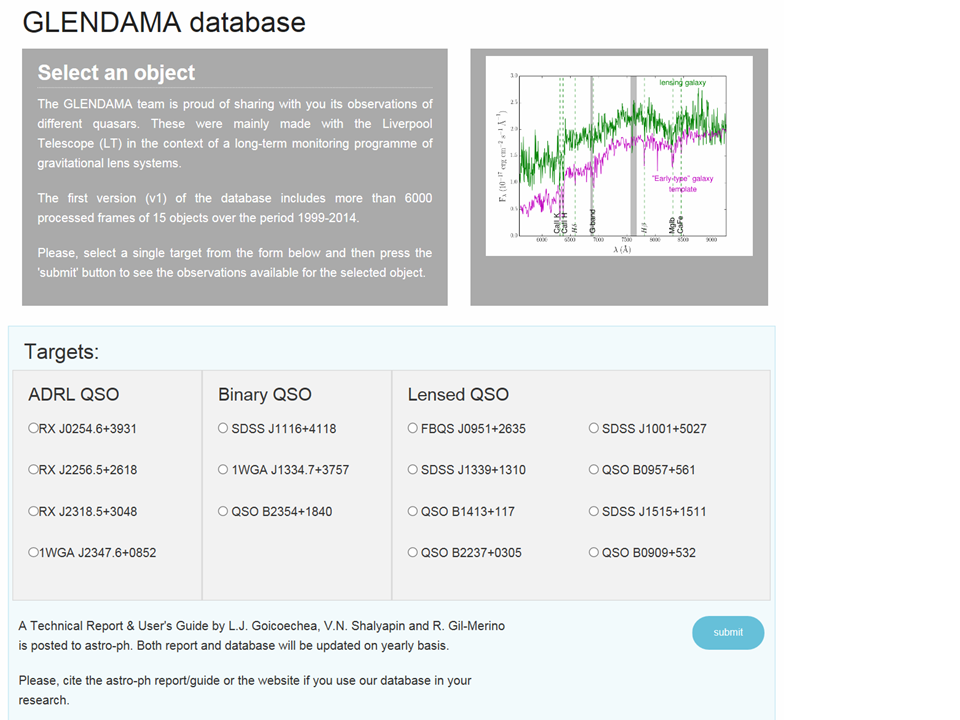}
\caption{Select an object. This is the first step in 
using the WUI for the GLENDAMA archive.\label{fig-1}}
\end{figure}

\begin{figure}
\includegraphics[width=1.3\textwidth]{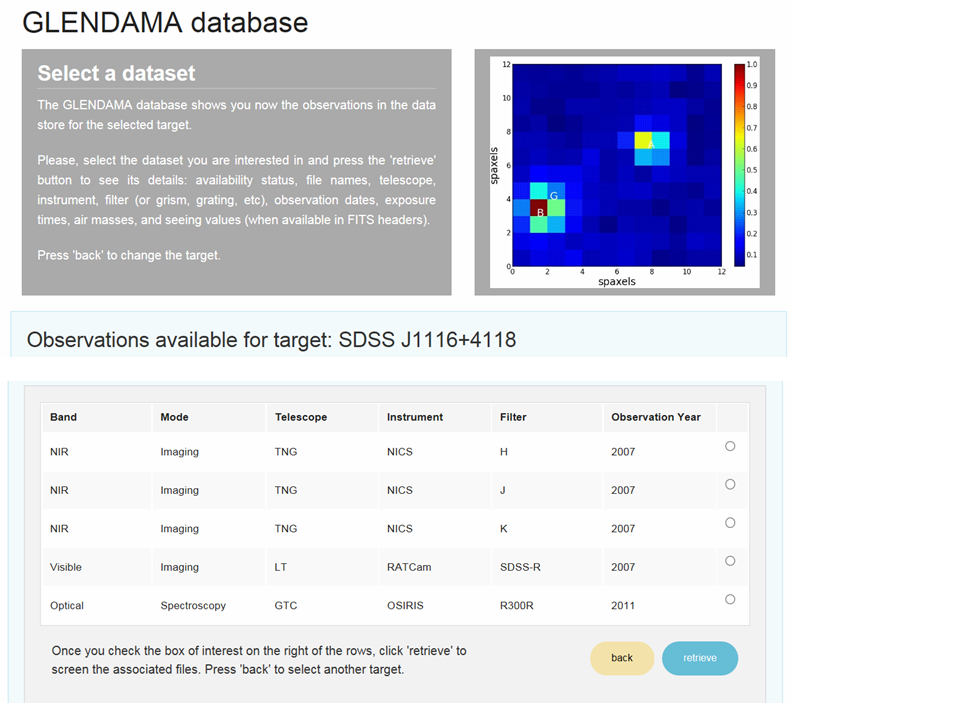}
\caption{Select a dataset. This is the second step in 
using the WUI for the GLENDAMA archive.\label{fig-2}}
\end{figure}

The web user interface (WUI) at \url{http://grupos.unican.es/glendama/database} 
allows users to surf the new archive, see all its content and freely download a 
significant fraction of data (see Table~\ref{tbl-2}). This interface is a 
3-step tool, where the first step is to {\bf select an object}. In 
Fig.~\ref{fig-1}, we display what a potential user would see initially. After 
selecting an object (see Table~\ref{tbl-1}) and clicking the submit button, the 
WUI shows the datasets available for the target of interest. In this second 
screen (see, e.g., Fig.~\ref{fig-2}), it is possible to {\bf select a dataset} 
and then to press the retrieve button to get its details. Apart from the 
telescope, the instrument and the filter (or grism, grating...), these details 
include the availability status (access), file names, observation dates, 
exposure times, air masses and seeing values (when available in FITS headers). 

\begin{figure}
\includegraphics[width=1.3\textwidth]{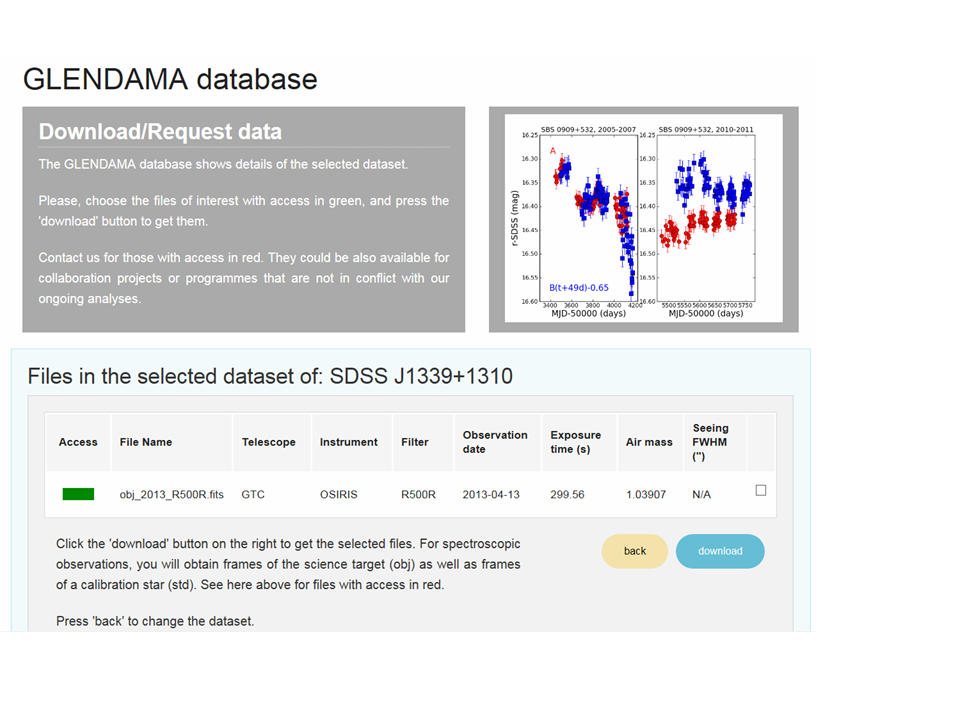}
\caption{Download data. The access is green for the 
selected dataset, and thus, the user is allowed to 
download it.\label{fig-3}}
\end{figure}

\begin{figure}
\includegraphics[width=1.3\textwidth]{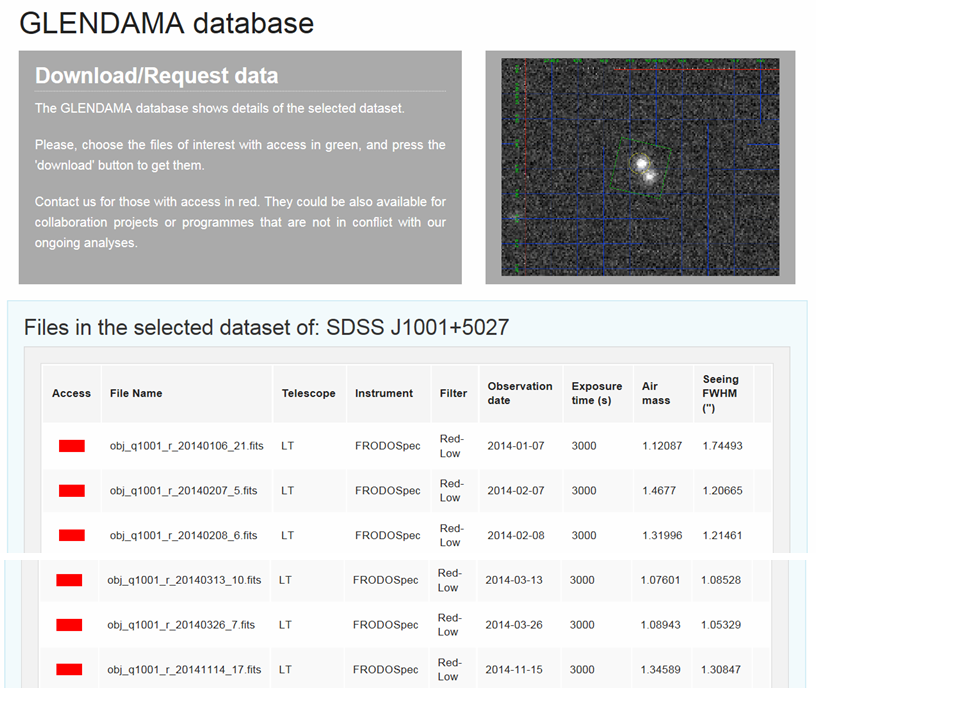}
\caption{Request data. The access is red for the 
selected dataset, so the user cannot download the 
associated files. However, under certain conditions, 
it is possible to request and get these "red files" 
(see main text).\label{fig-4}}
\end{figure}

In Fig.~\ref{fig-3} and Fig.~\ref{fig-4}, we show two screen examples for the 
third step, when the user can {\bf download or request data}. If the files are
freely available (access in green; see Fig.~\ref{fig-3}), the user is allowed 
to choose a number of files and download them. This is the last step. 
Alternatively, if the access is red (see Fig.~\ref{fig-4}), the observations 
could be also available for collaboration projects or programmes that are not 
in conflict with our ongoing analyses (see Table~\ref{tbl-2}). We are open to
suggestions on collaboration and new ideas to mine the "red archive". To 
contact us, please write to \email{goicol@unican.es} or \email{vshal@ukr.net}. 
We also note that files with current access in red will become freely available 
(green access) in successive versions of the GLENDAMA archive, just when the 
corresponding analyses are completed. 

Two remarks: first, a warning on the seeing values. These come from FITS
headers, and have an orientative character. For example, the FWHM of the seeing 
disc for RATCam, IOO and RINGO observations is directly estimated from frames.
Thus, it is a reliable reference. However, seeing values for FRODOSpec data are 
estimated before spectroscopic exposures, so these FWHMs may appreciably differ 
from true ones. Second, for spectroscopic observations, we offer frames of the 
science target and a calibration star. These FITS files for the main target and 
the star have labels including the expressions 'obj' and 'std', respectively. 

\section{Final comments and acknowledgements}

The first version of the GLENDAMA database consists of a datastore with $\sim$ 
30 GB of size, whose content is organised and visualised by using MySQL/PHP 
software. The MySQL/PHP tools were designed by RGM, and we thank the 
Universidad de Cantabria (UC) web service for making it possible the launch of 
this database. The astronomical frames in the datastore are part of a variety 
of observing programmes (see Table~\ref{tbl-2}). We acknowledge A. Ull\'an for 
doing observations with the Telescopio Nazionale Galileo (TNG) and processing 
raw frames of the Gravitational LensES International Time Project. We also have
an $\sim$ 20 GB complementary storage containing secondary frames.

We are indebted to C.J. Davis, J. Marchant, C. Moss and R.J. Smith for guidance 
in the preparation of the robotic monitoring project with the Liverpool 
Telescope (LT). We also acknowledge the staff of the LT for their development 
of the Phase 2 User Interface (which allows users to specify in detail the 
observations they wish the LT to make) and data reduction pipelines. The LT is 
operated on the island of La Palma by Liverpool John Moores University in the 
Spanish Observatorio del Roque de los Muchachos (ORM) of the Instituto de 
Astrof\'isica de Canarias (IAC) with financial support from the UK Science and 
Technology Facilities Council. 

We thank the support astronomers and other staff of the IAC and the ORM (J.A. 
Acosta, C. Alvarez, T. Augusteijn, R. Barrena, A. Cabrera, R.J. C\'ardenes R. 
Corradi, J. García, J. M\'endez, P. Monta\~{n}\'es, T. Pursimo, R. Rutten and 
C. Zurita, among others) for kind interactions regarding several observing 
programmes at the ORM. Based on observations made with the Gran Telescopio 
Canarias, installed at the Spanish ORM of the IAC, in the island of La Palma. 
This archive is also based on observations made with the Isaac Newton Group of 
Telescopes (Isaac Newton and William Herschel Telescopes), the Nordic Optical 
Telescope and the Italian TNG, operated on the island of La Palma by the Isaac 
Newton Group, the Nordic Optical Telescope Scientific Association and the 
Fundaci\'on Galileo Galilei of the Istituto Nazionale di Astrofisica, 
respectively, in the Spanish ORM of the IAC. 

We also use frames taken at the Observatorio del Teide (OT). We thank T. 
Granzer and A. Oscoz for managing and providing OT data. This archive is 
partially based on observations made with the IAC80 and STELLA 1 Telescopes 
operated on the island of Tenerife by the IAC and the AIP in the Spanish OT. 

We also thank the staff of the Chandra X-ray Observatory (CXO; E. Kellogg, H. 
Tananbaum and S.J. Wolk) and Swift Multi-wavelength Observatory (SMO; M. 
Chester and N. Gehrels) for their support during the preparation and execution 
of the monitoring campaign of QSO B0957+561 in 2010. The CXO Center is operated 
by the Smithsonian Astrophysical Observatory for and on behalf of the National 
Aeronautics Space Administration (NASA) under contract NAS803060. The SMO is 
supported at Penn State University by NASA contract NAS5-00136. 

The construction of the GLENDAMA database has been supported by the GLENDAMA 
project and a complementary action (PB97-0220-C02, AYA2000-2111-E, 
AYA2001-1647-C02-02, AYA2004-08243-C03-02, AYA2007-67342-C03-02, 
AYA2010-21741-C03-03 and AYA2013-47744-C3-2-P), all them financed by Spanish 
Departments of Education, Science, Technology and Innovation. RGM acknowledges 
grants of the AYA2010-21741-C03-03 and AYA2013-47744-C3-2-P projects to develop 
the core software of the database. This archive has been also possible thanks 
to the support of the UC. 






\clearpage

\clearpage

\clearpage

\clearpage

\clearpage

\begin{landscape}
\begin{deluxetable}{lcccccccccc}
\tabletypesize{\scriptsize}
\tablecaption{Basic Information for Objects in the GLENDAMA 
Database\label{tbl-1}}
\tablewidth{0pt}
\tablehead{
\colhead{Object}         					& 
\colhead{Class\tablenotemark{a}}       			&
\colhead{$z$\tablenotemark{b}}      			& 
\colhead{N$_{\rm ima}$\tablenotemark{c}}			&
\colhead{$\Delta \theta$\tablenotemark{d} (\arcsec)}	& 
\colhead{$r$\tablenotemark{e} (mag)}    			&
\colhead{ref}  							& 
\colhead{$\Delta t$\tablenotemark{f} (days)}  		&
\colhead{ref} 							& 
\colhead{Lensing Galaxy\tablenotemark{g}}  		&
\colhead{ref}}
\startdata
\object{RX J0254.6+3931} & ADRLQSO & 0.29 & 1 & \nodata & 15.4    & 1   & 
\nodata & \nodata & \nodata & \nodata \\

\object{QSO B0909+532}   & LQSO    & 1.38 & 2 & 1.1     & 16$-$17 & 2	& 
50 $^{+2}_{-4}$ & 3 & E ($z$ = 0.83) & 4 \\

\object{FBQS J0951+2635} & LQSO    & 1.24 & 2 & 1.1     & 17$-$18	& 5	& 
16 $\pm$ 2 & 6 & E ($z$ = 0.26) & 7 \\

\object{QSO B0957+561}   & LQSO    & 1.41 & 2 & 6.1     & 17	& 8   &
416.5 $\pm$ 1.0\tablenotemark{\star} & 9 & E-cD ($z$ = 0.36) & 10 \\

\object{SDSS J1001+5027} & LQSO    & 1.84 & 2 & 2.9   & 17.5$-$18 & 11	& 
119.3 $\pm$ 3.3 & 12 & E ($z$ = 0.41) & 13 \\

\object{SDSS J1116+4118} & BQSO & $\sim$ 3 & 2 & 13.8 & 18$-$19   & 14  & 
\nodata & \nodata & \nodata & \nodata \\

\object{1WGA J1334.7+3757} & BQSO & $\sim$ 1.89 & 2 & 82.2 & 20$-$20.5  & 
15  & \nodata & \nodata & \nodata & \nodata \\

\object{SDSS J1339+1310} & LQSO    &  2.24 & 2 &  1.7    & 19	& 16	& 
$\sim$ 40$-$50\tablenotemark{\star\star} & 17 & E ($z$ = 0.61) & 17 \\

\object{QSO B1413+117} & LQSO    &  2.55 & 4 &  1.4    & $\sim$ 18 & 18	& 
23 $\pm$ 4 & 19 & ? ($z$ = 1.88)\tablenotemark{\dagger} & 19 \\

\object{SDSS J1515+1511} & LQSO    &  2.05 & 2 &  2.0    & 18$-$19 & 20	& 
$\sim$ 200\tablenotemark{\star\star} & \nodata & 
S ($z$ = 0.74)\tablenotemark{\ddagger}  & 20 \\

\object{QSO B2237+0305} & LQSO    &  1.69 & 4 &  1.8    & 17.5$-$18.5	& 
21	& 1.5 $\pm$ 2.0 & 22 & SBb ($z$ = 0.04) & 21 \\

\object{RX J2256.5+2618} & ADRLQSO & 0.12 & 1 & \nodata & 16.7    & 1   & 
\nodata & \nodata & \nodata & \nodata \\

\object{RX J2318.5+3048} & ADRLQSO & 0.10 & 1 & \nodata & $\leq$ 17  	& 
1   & \nodata & \nodata & \nodata & \nodata \\

\object{1WGA J2347.6+0852} & ADRLQSO & 0.29 & 1 & \nodata & 16.2  & 1   & 
\nodata & \nodata & \nodata & \nodata \\

\object{QSO B2354+1840} & BQSO & $\sim$ 1.68 & 2 & 96.2 & $\leq$ 19     & 
23  & \nodata & \nodata & \nodata & \nodata \\
\enddata
\tablenotetext{a}{Class of object: LQSO $\equiv$ Lensed QSO, BQSO $\equiv$ 
Binary QSO, and ADRLQSO $\equiv$ Accretion-Dominated Radio-Loud QSO.}
\tablenotetext{b}{Redshift.}
\tablenotetext{c}{Number of QSO images.}
\tablenotetext{d}{Angular separation between images for double/binary QSOs, 
or typical angular size for quadruple QSOs.}
\tablenotetext{e}{$r$-band magnitudes of QSO images. These values should be 
interpreted with caution, since some data are for red filters different to 
$r$-Sloan, and we deal with variable objects.}
\tablenotetext{f}{Measured time delay for double QSOs, or the longest of the
measured delays for quadruple QSOs (1$\sigma$ confidence interval). When a 
direct measure is not available, the expected value is given.}
\tablenotetext{g}{Classification (redshift).}
\tablenotetext{\star}{Time delay in the $g$ band. There is evidence of 
chromaticity in the optical delay, since it is 420.6 $\pm$ 1.9 days in the 
$r$ band.}
\tablenotetext{\star\star}{Time delay predicted by the lens model.}
\tablenotetext{\dagger}{Redshift from gravitational lensing data and a 
concordance cosmology. This measurement is in reasonable agreement with the 
photometric redshift of the secondary lensing galaxy and the most distant 
overdensity, as well as the redshift of one of the absorption systems.}
\end{deluxetable}
\end{landscape}

\clearpage

\begin{landscape}
\begin{scriptsize}
$^{\ddagger}$ Redshift based on photometric data and the 
strong absorber in the spectrum of the faintest quasar image.

References: [1] \citet{Lan08}, and references therein $-$ 
[2] \citet{Koc97} $-$ 
[3] \citet{Hai13} \citep[see also][]{Goi08} $-$ 
[4] \citet{Osc97,Leh00,Lub00} $-$ 
[5] \citet{Sch98} $-$ 
[6] \citet{Jak05} $-$ 
[7] \citet{Koc00,Eig07} $-$
[8] \citet{Wal79,Wey79} $-$
[9] \citet{Sha12} \citep[see also][]{Kun97,Sha08} $-$
[10] \citet{Sto80,You80,You81} $-$
[11] \citet{Ogu05} $-$
[12] \citet{Rat13} $-$ 
[13] \citet{Ina12} $-$
[14] \citet{Hen06,Hen10} $-$
[15] \citet{Mch98,Zhd01} $-$
[16] \citet{Ina09} $-$
[17] \citet{Sha14b} $-$
[18] \citet{Mag88} $-$
[19] \citet{Goi10} $-$
[20] \citet{Ina14} $-$
[21] \citet{Huc85,Yee88} $-$
[22] \citet{Vak06} $-$
[23] \citet{Bor96,Zhd01}.
\end{scriptsize}
\end{landscape}

\clearpage

\begin{landscape}
\begin{deluxetable}{lcccccccccc} 
\tablecolumns{10} 
\tabletypesize{\scriptsize}
\tablecaption{Datasets and Current Status\label{tbl-2}}
\tablewidth{0pt} 
\tablehead{ 
\colhead{}    						&  
\multicolumn{2}{c}{Imaging} 				&   
\colhead{}   						& 
\colhead{Spectroscopy} 					&
\colhead{}   						& 
\colhead{Polarimetry}					&
\colhead{}   						& 
\colhead{}   						& 
\colhead{}   						\\ 
\cline{2-3} 						\\ 
\colhead{Obs. Period} 					& 
\colhead{NUV/Visible}    				&  
\colhead{NIR} 						& 
\colhead{}   						&
\colhead{Optical}    					& 
\colhead{}   						& 
\colhead{Optical}    					& 
\colhead{N$_{\rm frames}$}   				& 
\colhead{Size (GB)}    					& 
\colhead{Status}}
\startdata 
\cutinhead{RX J0254.6+3931} 
2011 Sep & LT/RATCam/$gr$ & \nodata & & LT/FRODOSpec/BR & & 
\nodata & 12 & 0.18 & F\tablenotemark{1} \nl
2011$-$2012 & LT/RATCam/$g$ & LT/RATCam/$i$ & & \nodata & & 
\nodata & 88 & 0.18 & F\tablenotemark{2} \nl
\cutinhead{QSO B0909+532} 
2005$-$2007 & LT/RATCam/$gr$ & \nodata & & \nodata & & 
\nodata & 451 & 0.92 & F\tablenotemark{3} \nl
2010$-$2012 & LT/RATCam/$r$ & \nodata & & \nodata & & 
\nodata & 119 & 0.25 & F\tablenotemark{3} \nl
2012$-$2014 & LT/IOO/$gr$ & LT/IOO/$i$\tablenotemark{a} & 
& \nodata & & \nodata & 279 & 2.25 & R\tablenotemark{4} \nl
\cutinhead{FBQS J0951+2635} 
2007 Feb-May & \nodata & LT/RATCam/$i$ & & \nodata & & 
\nodata & 259 & 0.52 & F\tablenotemark{5} \nl
2009$-$2012 & LT/RATCam/$r$ & \nodata & & \nodata & & 
\nodata & 29 & 0.06 & R\tablenotemark{6} \nl
2013$-$2014 & LT/IOO/$r$ & \nodata & & \nodata & & 
\nodata & 27 & 0.22 & R\tablenotemark{6} \nl
\cutinhead{QSO B0957+561} 
1999$-$2005 & IAC80/CCD/$BVR$\tablenotemark{b} & 
IAC80/CCD/$I$\tablenotemark{b} & & \nodata & & \nodata & 1108 & 
4.13 & R\tablenotemark{7} \nl
2000 Feb-Mar & NOT/StanCam/$VR$ & \nodata & & \nodata & & 
\nodata & 77 & 0.25 & F\tablenotemark{8} \nl
2005$-$2014 & LT/RATCam/$gr$\tablenotemark{c} & \nodata & 
& \nodata & & \nodata & 1192 & 2.34 & F/R\tablenotemark{9} \nl
2007 Dec & \nodata & TNG/NICS/$JHK$\tablenotemark{d} & & 
\nodata & & \nodata & 3 & 0.01 & F\tablenotemark{10} \nl
2008 Mar & \nodata & \nodata & & INT/IDS/R300V & & 
\nodata & 3 & $<$ 0.01 & R\tablenotemark{11} \nl
2009$-$2013 & \nodata & \nodata & & NOT/ALFOSC/G7G14 & & 
\nodata & 14 & 0.02 & R\tablenotemark{11} \nl
2010 Jan-Jun & UVOT/MIC/$U$ & LT/RATCam/$iz$\tablenotemark{e} 
& & \nodata & & \nodata & 154 & 0.48 & F\tablenotemark{12} \nl
2010$-$2014 & \nodata & \nodata & & 
LT/FRODOSpec/BR\tablenotemark{f} & & \nodata & 228 & 4.76 & 
R\tablenotemark{13} \nl
2011$-$2012 & \nodata & \nodata & & \nodata & & 
LT/RINGO/V+R & 32 & 0.02 & F\tablenotemark{14} \nl
2012$-$2014 & LT/IOO/$gr$ & \nodata & & \nodata & & \nodata 
& 116 & 0.93 & R\tablenotemark{15} \nl
2013$-$2014 & \nodata & \nodata & & \nodata & & 
LT/RINGO/BGR & 216 & 2.57 & R\tablenotemark{14} \\
\cline{1-11} \\
& & & & & & & & & & \nl
\cutinhead{SDSS J1001+5027} 
2010 Feb-May & LT/RATCam/$g$ & \nodata & & \nodata & & 
\nodata & 46 & 0.09 & F\tablenotemark{16} \nl
2013$-$2014 & LT/IOO/$gr$\tablenotemark{g} & \nodata & & 
LT/FRODOSpec/BR & & \nodata & 68 & 1.03 & R\tablenotemark{17} \nl
2014 Jan/Mar & \nodata & \nodata & & \nodata & & 
LT/RINGO/BGR & 72 & 1.52 & R\tablenotemark{14} \nl
\cutinhead{SDSS J1116+4118} 
2007 May-Jul & LT/RATCam/$r$ & \nodata & & 
\nodata & & \nodata & 10 & 0.02 & F\tablenotemark{18} \nl
2007 Dec & \nodata & TNG/NICS/$JHK$\tablenotemark{d} & & 
\nodata & & \nodata & 3 & 0.01 & R\tablenotemark{19} \nl
2011 Mar-Apr & \nodata & \nodata & & GTC/OSIRIS/R300R & & 
\nodata & 5 & $<$ 0.01 & F\tablenotemark{20} \nl
\cutinhead{1WGA J1334.7+3757} 
2006 Jan & NOT/ALFOSC/$R$\tablenotemark{d} & 
NOT/ALFOSC/$I$\tablenotemark{d} & & \nodata & & \nodata & 2 & 0.04 
& R\tablenotemark{21} \nl
2007 Apr & \nodata & \nodata & & WHT/ISIS/BR & & \nodata & 11 & 
0.07 & R\tablenotemark{22} \nl
\cutinhead{SDSS J1339+1310} 
2009/2012 & LT/RATCam/$r$ & \nodata & & \nodata & & 
\nodata & 293 & 0.59 & R\tablenotemark{23} \nl
2010 Jun-Jul & \nodata & LT/RATCam/$i$ & & \nodata & & 
\nodata & 20 & 0.04 & F\tablenotemark{24} \nl
& & NOT/ALFOSC/$I$\tablenotemark{d} & & & & & 1 & $<$ 0.01 & 
F\tablenotemark{24} \nl
2013 Apr & \nodata & \nodata & & GTC/OSIRIS/R500R & & 
\nodata & 2 & $<$ 0.01 & F\tablenotemark{25} \nl
2013$-$2014 & LT/IOO/$r$ & \nodata & & \nodata & & 
\nodata & 106 & 0.85 & R\tablenotemark{23} \nl
2013$-$2014 & \nodata & \nodata & & \nodata & & 
LT/RINGO/BGR & 72 & 0.53 & R\tablenotemark{14} \nl
2014 Mar & \nodata & \nodata & & GTC/OSIRIS/R500R & & 
\nodata & 4 & $<$ 0.01 & R\tablenotemark{25} \nl
2014 May & \nodata & \nodata & & GTC/OSIRIS/R500B & & 
\nodata & 4 & $<$ 0.01 & R\tablenotemark{25} \nl
\cutinhead{QSO B1413+117} 
2008 Feb-Jul & LT/RATCam/$r$ & \nodata & & \nodata & & 
\nodata & 61 & 0.12 & F\tablenotemark{26} \nl
2011 Mar/Jun & \nodata & \nodata & & GTC/OSIRIS/R300R & & 
\nodata & 8 & $<$ 0.01 & F\tablenotemark{20} \nl
2013$-$2014 & LT/IOO/$r$ & \nodata & & \nodata & & 
\nodata & 58 & 0.47 & R\tablenotemark{27} \\
\cutinhead{SDSS J1515+1511} 
2014 Feb-Aug & LT/IOO/$r$ & \nodata & & \nodata & & 
\nodata & 114 & 0.92 & R\tablenotemark{28} \\
\cutinhead{QSO B2237+0305} 
2007$-$2009 & LT/RATCam/$gr$\tablenotemark{h} & \nodata & & 
\nodata & & \nodata & 174 & 0.35 & R\tablenotemark{29} \nl
2013 Jun-Dec & \nodata & \nodata & & LT/FRODOSpec/BR & & 
LT/RINGO/BGR & 216 & 1.10 & R\tablenotemark{30} \nl
2013$-$2014 & LT/IOO/$gr$ & \nodata & & \nodata & & 
\nodata & 94 & 0.76 & R\tablenotemark{29} \\
\cutinhead{RX J2256.5+2618} 
2012 Jul-Dec & LT/RATCam/$gr$ & \nodata & & \nodata & & 
\nodata & 86 & 0.17 & F\tablenotemark{31} \nl
\cutinhead{RX J2318.5+3048} 
2012 Jul-Dec & LT/RATCam/$gr$ & \nodata & & \nodata & & 
\nodata & 96 & 0.19 & F\tablenotemark{31} \nl
\cutinhead{1WGA J2347.6+0852} 
2011$-$2012 & LT/RATCam/$g$ & LT/RATCam/$i$ & & \nodata & & 
\nodata & 58 & 0.12 & F\tablenotemark{2} \nl
2013 Jun-Jul & STELLA/WiFSIP/$UBgVr$ & \nodata & & \nodata & & 
\nodata & 190 & 1.81 & F\tablenotemark{32} \nl
\cutinhead{QSO B2354+1840} 
2007 Dec & \nodata & TNG/NICS/$JHK$\tablenotemark{d} & & 
\nodata & & \nodata & 3 & 0.01 & R\tablenotemark{19} \nl
\enddata 
\tablecomments{Table \ref{tbl-2} does not include our X-ray (0.1$-$10 keV)
monitoring campaign of QSO B0957+561 during the first semester of 
2010. This was performed with Chandra/HRMA/ACIS-S3 [Freely available, 
Programme: DDT\# 10708333, Data analysis: \citet{Gil12}]. We use three 
abbreviations A/B/C to describe the facilities, where A $\equiv$ Telescope, 
B $\equiv$ Instrument, and C $\equiv$ Details (filters, spectral arms, 
grisms or gratings). We explain what all} 
\end{deluxetable} 
\end{landscape}

\clearpage

\begin{landscape}
\begin{scriptsize} 
\noindent
these abbreviations mean in 
section~\ref{objfac}. In the last column, we also comment on the dataset 
availability status: F $\equiv$ Free and R $\equiv$ (upon) Request (see 
section~\ref{wui} for details on obtaining R data), as well as the original 
programme, and the completed (COMP), ongoing (ONGO), planned (PLAN) and 
published (PUBL) GLENDAMA analyses.\\
$^{a}$ No data in 2014.\\
$^{b}$ Poorer sampling in the $BI$ bands.\\
$^{c}$ Only a few frames in 2013$-$2014.\\
$^{d}$ Combined frames (deep imaging observations).\\
$^{e}$ No frames in 2010 January at NIR wavelengths, and several 
frames in late 2010 and early 2011 in the $i$ band.\\
$^{f}$ Poorer sampling in 2010, 2013 and 2014.\\
$^{g}$ $\sim$ 90\% of frames in the $r$ band.\\
$^{h}$ No data in 2007$-$2008 in the $g$-band.\\
$^{1}$ Programme: XCL04BL2.\\
$^{2}$ Programme: XCL04BL2, Data analysis: COMP $-$ Light curves 
in the $gi$ bands.\\
$^{3}$ Programme: XCL04BL2, Data analysis: PUBL $-$
\citet{Goi08,Hai13}.\\
$^{4}$ Programme: XCL04BL2, Data analysis: COMP $-$ Light curves 
in the $gri$ bands; ONGO $-$ Multi-wavelength extrinsic variations vs 
microlensing simulations.\\
$^{5}$ Programme: XCL04BL2, Data analysis: PUBL $-$ 
\citet{Sha09}.\\
$^{6}$ Programme: XCL04BL2, Data analysis: PLAN $-$ Time 
evolution of red fluxes after 2007.\\
$^{7}$ Programme: IAC Gravitational Lenses Monitoring, Data 
analysis: COMP $-$ Light curves in the $VR$ bands; ONGO $-$ Astrophysical 
scenarios to explain the observed long-term variability.\\
$^{8}$ Programme: Gravitational Lenses International Time 
Project (GLITP), Data analysis: PUBL $-$ \citet{Ull03}.\\
$^{9}$ F($\leq$ 2010) \& R($>$ 2010), Programme: XCL04BL2, Data 
analysis: PUBL $-$ \citet{Sha08,Goi09,Gil12,Sha12}; ONGO $-$ Astrophysical 
scenarios to explain the observed long-term variability.\\
$^{10}$ Programme: A16CAT128.\\
$^{11}$ Programmes: CAT-ST \& NOT-SP 40$-$401, Data analysis: 
COMP $-$ Spectrophotometric monitoring.\\
$^{12}$ Programmes: Swift/TOO\# 31567 \& XCL04BL2, Data 
analysis: PUBL $-$ \citet{Gil12,Goi12a}.\\
$^{13}$ Programme: XCL04BL2, Data analysis: PUBL $-$ 
\citet{Sha14a}; COMP $-$ Spectrophotometric monitoring.\\
$^{14}$ Programme: XCL04BL2, Data analysis: ONGO $-$ 
Polarimetric follow-up study.\\
$^{15}$ Programme: XCL04BL2, Data analysis: ONGO $-$ 
Astrophysical scenarios to explain the observed long-term variability.\\
$^{16}$ Programme: XCL04BL2, Data analysis: COMP $-$ Light 
curves in the $g$ band.\\
$^{17}$ Programme: XCL04BL2, Data analysis: PLAN $-$ Time 
evolution of the flux ratio for continuum and line emitting regions.\\
$^{18}$ Programme: XCL04BL2, Data analysis: COMP $-$ Variability
in the $r$ band.\\
$^{19}$ Programme: A16CAT128, Data analysis: ONGO 
$-$ Understanding widely separated pairs of quasars at similar redshift.\\
$^{20}$ Programme: GTC35$-$11A, Data analysis: PUBL $-$ 
\citet{Sha13}; ONGO $-$ Understanding widely separated pairs of quasars 
at similar redshift.\\
$^{21}$ Programme: CAT-ST, Data analysis: ONGO 
$-$ Understanding widely separated pairs of quasars at similar redshift.\\
$^{22}$ Programme: ING-SP SW2007a51, Data analysis: ONGO 
$-$ Understanding widely separated pairs of quasars at similar redshift.\\
$^{23}$ Programme: XCL04BL2, Data analysis: PUBL $-$ 
\citet{Goi12b}; COMP $-$ Light curves in the $r$ band; ONGO $-$ Detailed 
study in the time and wavelength domains: time delay, microlensing events, 
extinction, macrolens flux ratio, microlensing spectrum, etc.\\
$^{24}$ Programmes: XCL04BL2 \& CAT-ST, Data analysis: PUBL $-$
\citet{Sha14b}.\\
$^{25}$ Programmes: GTC30$-$13A \& GTC82$-$14A, Data analysis: 
PUBL $-$ \citet{Sha14b}; ONGO $-$ Detailed study in the time and wavelength 
domains: time delay, microlensing events, extinction, macrolens flux ratio,
microlensing spectrum, etc.\\
$^{26}$ Programme: XCL04BL2, Data analysis: PUBL $-$ 
\citet{Goi10}.\\
$^{27}$ Programme: XCL04BL2, Data analysis: COMP $-$ Light 
curves in the $r$ band; PLAN $-$ Improving delay estimates and detecting 
microlensing variability.\\
$^{28}$ Programme: XCL04BL2, Data analysis: ONGO $-$ Light 
curves and time delay.\\
$^{29}$ Programme: XCL04BL2, Data analysis: COMP $-$ Light 
curves; ONGO $-$ Following the variability of the four QSO images.\\
$^{30}$ Programme: XCL04BL2, Data analysis: ONGO $-$ Checking 
the performance of FRODOSpec and RINGO.\\
$^{31}$ Programme: XCL04BL2, Data analysis: COMP $-$ Light 
curves in the $gr$ bands.\\
$^{32}$ Programme: 42-Stella4$-$13A, Data analysis: COMP $-$ 
Checking the performance of WiFSIP.
\end{scriptsize}
\end{landscape} 







\end{document}